\documentclass[10pt,twocolumn]{article}

\PassOptionsToPackage{table}{xcolor}

\usepackage{fardlab}

\usepackage{textcomp}
\usepackage{pifont}
\usepackage{listings}
\usepackage{algorithm}
\usepackage{algpseudocode}
\usepackage{xurl}
\usepackage{fontawesome5}

\fancypagestyle{fard@firstpage}{%
  \fancyhf{}%
  \fancyfoot[C]{\small\color{fardMuted}\thepage}%
}

\newcommand{\sysname}{ARISE}
\newcommand{\bench}{SWE-bench Lite}
\newcommand{\backbone}{Qwen2.5-Coder-32B-Instruct}

\newcommand{\condtext}{\textsc{RAG}}
\newcommand{\condstruct}{\textsc{ARISE-Structural}}
\newcommand{\condslice}{\textsc{ARISE-Slicing}}
\newcommand{\condfull}{\textsc{ARISE-Full}}
\newcommand{\condcoarse}{\textsc{ARISE-Coarse}}

\newcommand{\searchentities}{\texttt{search\_entities}}
\newcommand{\traverserelations}{\texttt{traverse\_relations}}
\newcommand{\getenclosing}{\texttt{get\_enclosing\_scopes}}
\newcommand{\getcodespan}{\texttt{get\_code\_span}}
\newcommand{\getentityinfo}{\texttt{get\_entity\_info}}
\newcommand{\getslice}{\texttt{get\_dataflow\_slice}}
\newcommand{\buildcontext}{\texttt{build\_context\_bundle}}
\newcommand{\ranksuspect}{\texttt{rank\_suspect\_regions}}

\title{ARISE: A Repository-level Graph Representation and Toolset for Agentic Program Repair and Fault Localization}

\shorttitle{ARISE: Graph Representation and Toolset for Agentic Repair}

\author{
  Shahd Seddik\textsuperscript{1,*}\enspace
  Fahd Seddik\textsuperscript{1}\enspace
  Amirrezza Esmaeili\textsuperscript{1}\enspace
  Mahdieh Sadatbenis\textsuperscript{1}\enspace
  Fatemeh Fard\textsuperscript{1}
}

\affil{1}{FARD Lab, University of British Columbia, Kelowna, Canada}

\authornote{*}{Corresponding author: \href{mailto:shahd.seddik@ubc.ca}{shahd.seddik@ubc.ca}}

\institutionlogos{\normalsize\bfseries\color{fardTeal}\faGlobe~\href{https://fard-lab.github.io/ARISE/}{Project page: fard-lab.github.io/ARISE}}

\begin{document}

\twocolumn[
  \maketitle
  \begin{abstract}
    Automated program repair at repository scale requires an agent to locate a fault among thousands of files and synthesize a correct patch.
    Existing graph-based agents represent how a repository is organized into files, classes, and functions, but they do not model how variable values flow within a procedure, which leaves the agent without the semantic precision that function-level and line-level localization demand.
    We present \sysname{} (\textbf{A}gentic \textbf{R}epository-level \textbf{I}ssue \textbf{S}olving \textbf{E}ngine), a framework-agnostic toolset that builds a multi-granularity program graph, extending structural relationships down to statement-level nodes connected by intra-procedural definition-use edges, and exposes it through a three-tier tool API that mounts on any tool-use agentic framework.
    The central primitive is data-flow slicing, a queryable agent tool that traces in a single call which statements define or consume a variable of interest.
    On SWE-bench Lite (300 real GitHub issues across 11 Python repositories) with the open-source Qwen2.5-Coder-32B-Instruct backbone, mounting \sysname{} on SWE-agent as the host resolves 22.0\% of issues (66/300), a 4.7 percentage-point gain over the unmodified SWE-agent baseline under the identical backbone and host.
    We show this gain is largely attributable to sharper localization, with Function Recall@1 (R@1) rising from 0.43 to 0.60 (a 40\% relative gain) and Line R@1 from 0.26 to 0.41 (a 58\% relative gain).
    Controlled ablations attribute the improvement to the data-flow graph rather than the tool schema, and we further mount the same toolset on a second host framework to study its portability.
    Decoupled from any single scaffold, the graph builder and slicing API form a drop-in toolset for future repair research.
  \end{abstract}
  \vspace{1.2em}
]
\thispagestyle{fard@firstpage}

\section{Introduction}
\label{sec:introduction}

Automated Program Repair (APR) aims to automatically generate source-code changes that fix software defects, and its first step is Fault Localization (FL), the task of identifying the specific files, functions, and lines responsible for a failure~\cite{hossain2024-bug-localization-survey}.
This step is expensive in practice: developers spend an estimated 35\% to 50\% of their working time on validation and debugging, and debugging, testing, and verification together account for 50\% to 75\% of total software-project budgets, amounting to more than 100 billion USD annually~\cite{odell2017debugging}, figures rooted in an industry study from the Cambridge Judge Business School~\cite{britton2013reversible} and echoed in recent surveys of Large Language Model (LLM)-based repair~\cite{zhang2025llm4apr-slr}.
The difficulty is amplified by scale, because modern software is developed in large repositories where code spans many files and packages with rich build and dependency relationships~\cite{potvin2016google}.
Operating at this scale requires an automated system to identify relevant code units across files, to follow call and data dependencies through the codebase, and to maintain architectural context.
Even strong LLM-based systems resolve only a minority of real repository-level issues when applied naively~\cite{jimenez2023swe}, and accurate localization is the decisive bottleneck that separates a successful repair from a failed one \cite{yu2025orcaloca}.

LLMs and LLM-based agents are now widely applied to FL and APR over large repositories~\cite{zhang2025llm4apr-slr}.
Two design philosophies dominate.
\emph{Agentic} systems interleave reasoning with tool actions, letting the model open, edit, and test code across a project.
SWE-agent introduces an Agent-Computer Interface for this purpose~\cite{sweagent2024neurips}, AutoCodeRover navigates the repository through structure-aware search APIs before patching~\cite{zhang2024autocoderover}, SpecRover extends AutoCodeRover with iterative specification inference and a reviewer agent for patch vetting~\cite{ruan2025specrover}, and OrcaLoca adds priority-based scheduling and context pruning for localization~\cite{yu2025orcaloca}.
\emph{Agentless} system instead fix a multi-phase pipeline of localization, generation, and validation, matching tool-driven agents at lower cost~\cite{xia2025agentless}, and SWE-Fixer trains open-source LLMs around a coarse-to-fine retrieval-and-edit pipeline~\cite{xie2025swefixer}.
A complementary line gives agents explicit code structure through graphs, including RepoGraph~\cite{repograph2025iclr}, LocAgent~\cite{chen2025locagent}, KGCompass~\cite{yang2025kgcompass}, and CodexGraph~\cite{liu2025codexgraph}, with CGM integrating structural code graphs directly into an open-source LLM~\cite{tao2025cgm}, which improve navigation and downstream reasoning over flat retrieval.

Despite this progress, current graph-based APR and FL systems share three limitations.
First, their graphs model structure only, namely files, classes, and functions, and stop above the statement level, so they cannot represent how a value is computed inside a procedure~\cite{repograph2025iclr, chen2025locagent, jiang2025cosil}.
Second, the graph is typically precomputed and consumed as a fixed feature source rather than a substrate the agent can query on demand during reasoning~\cite{graphcodebert2021}.
Third, none exposes a way to trace how a specific variable is defined and propagated within a function.
This matters because the faulty statement is frequently not the one named in the issue but one reachable from it only by following a definition-use chain.

We present \sysname{} (\textbf{A}gentic \textbf{R}epository-level \textbf{I}ssue \textbf{S}olving \textbf{E}ngine), a framework-agnostic toolset that targets this gap through a single design decision, namely augmenting a structural repository graph with statement-level nodes and intra-procedural definition-use edges, and exposing the resulting data-flow slices as a first-class tool primitive through a three-tier tool API.
\sysname{} is decoupled from any specific agentic scaffold and mounts on any host framework that exposes a tool-call interface, requiring only that the host inject the \sysname{} tool schema and dispatch tool calls across its turns.
For the controlled experiments we use SWE-agent~\cite{sweagent2024neurips} as the primary host, reusing its Agent-Computer Interface and multi-turn protocol, and we mount the same toolset on a second host framework to study portability.
This gives the backbone model precise, queryable evidence about both how the repository is organized and how values flow within the functions under suspicion.
The novelty is not slicing itself, which is a classical analysis, but exposing it as an on-demand agent primitive rather than a static preprocessing artifact, so that a single backward-slice query resolves in one step the def-use chain that a structure-only graph can only approximate through many traversal hops.

\begin{figure*}[t]
  \centering
  \includegraphics[width=\textwidth]{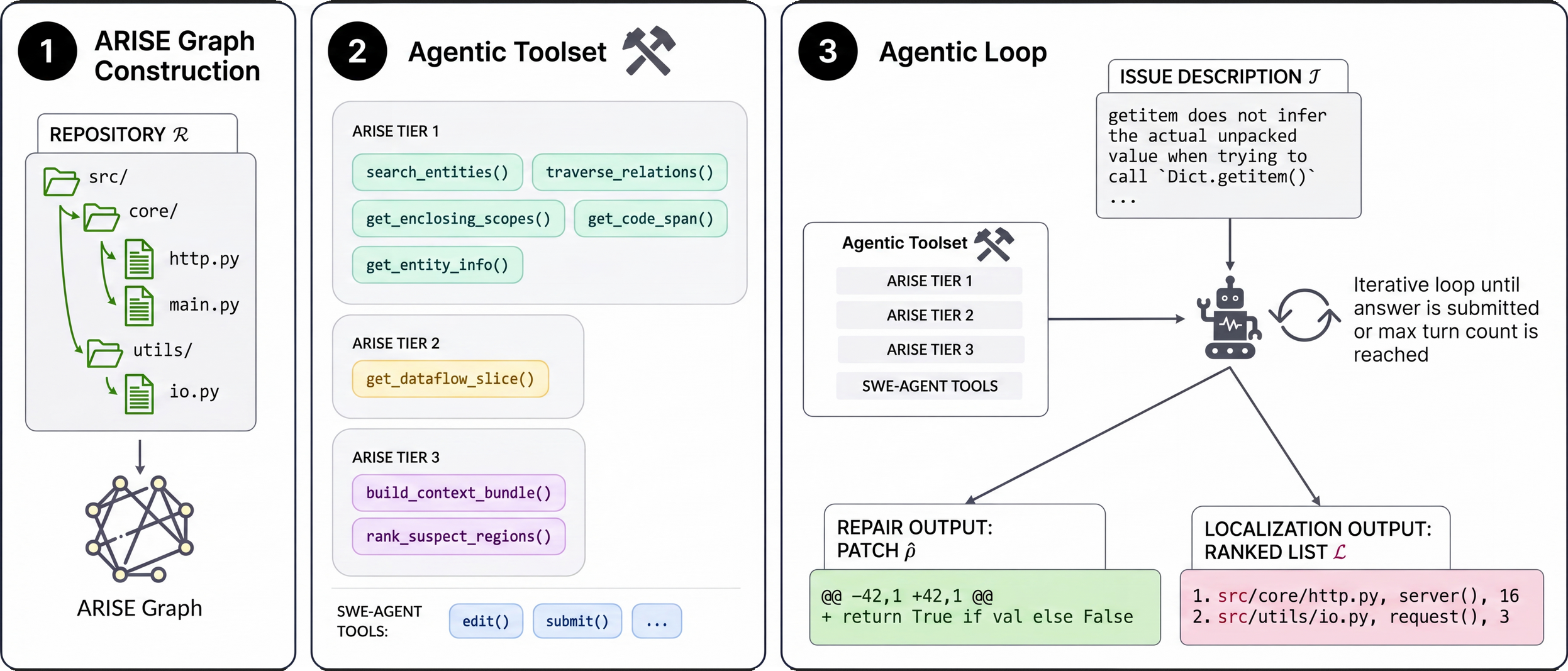}
  \caption{\sysname{} pipeline overview. \textbf{Phase~1:} given a repository snapshot, \sysname{} constructs a multi-granularity program graph that combines structural relationships with intra-procedural data-flow edges (definition-use chains at statement level). \textbf{Phase~2:} the agentic tool API adds three tiers, where Tier~1 provides structural navigation, Tier~2 adds data-flow slicing that traces how variables are defined and consumed, and Tier~3 assembles the collected evidence into a ranked, token-budgeted context bundle. \textbf{Phase~3:} an LLM-based agent interacts with the graph through this API and produces a ranked list of fault locations (localization) or a unified diff (repair).}
  \label{fig:overview}
\end{figure*}

We make three contributions.

\begin{enumerate}[leftmargin=*, label=(\arabic*)]
\item \textbf{Multi-granularity program graph.}
  A repository graph that unifies structural and data-flow information at statement granularity, letting an agent query not only how code is organised (files, classes, functions) but also how values are defined and propagated within a procedure, evidence unavailable in structure-only graphs (Section~\ref{sec:graph}).

\item \textbf{Data-flow slicing as an agent primitive.}
  We expose data-flow slicing as a first-class, queryable tool, allowing an agent to retrieve the precise set of statements that affect, or are affected by, a variable of interest, with backward, forward, and bidirectional queries returning structured output the model reasons over directly (Tier~2, Section~\ref{sec:methodology}).

\item \textbf{Portable, framework-agnostic toolset.}
  We design \sysname{}'s graph builder and slicing API as a self-contained component with a thin integration contract, so any tool-use-capable framework can use it without re-implementing the underlying analysis (Section~\ref{sec:generalization}).
\end{enumerate}

We organize the evaluation around three research questions.

\textbf{RQ1 (Repair).} \emph{Does \sysname{} improve end-to-end bug repair over text-only, agentic, and structural-graph baselines?}
Repair is the end goal, so we first measure whether the graph and slicing tools resolve more issues than the unmodified SWE-agent scaffold and a structural-graph configuration.

\textbf{RQ2 (Localization mechanism).} \emph{Is the repair improvement explained by better fault localization?}
We treat localization as the analytic lens on repair.
By measuring localization recall at file, function, and line granularity and its rank correlation with repair success, this question tests whether localization precision is the binding constraint that is improved by the slicing tools~\cite{hossain2024-bug-localization-survey, xu2025flexfl}.

\textbf{RQ3 (Ablation).} \emph{Which components of \sysname{} contribute to the gains, and how?}
\sysname{} introduces structural graph tools, data-flow slicing, and context bundling.
We isolate each through controlled conditions that add or remove one component while holding all else constant.

We evaluate on \bench{}~\cite{jimenez2023swe} with \backbone{}, comparing against a BM25 retrieval lower bound, the unmodified SWE-agent agentic baseline, and a structural-graph configuration that reproduces LocAgent and RepoGraph capabilities.
With SWE-agent as the host, \sysname{} resolves 22.0\% of issues (66/300), a 4.7 percentage-point gain over SWE-agent, and improves Function Recall@1 from 0.43 to 0.60 (a 40\% relative gain) and Line Recall@1 from 0.26 to 0.41 (a 58\% relative gain).
The absolute resolve rate is bounded by our use of an open-source backbone rather than a frontier closed model, so we report controlled within-setting deltas under a fixed host and backbone.
Data-flow slicing is the single largest component, lifting Function Recall@1 from 0.50 to 0.57 and Pass@1 by 2.0 percentage points, and controlled ablations attribute this gain to the data-flow graph itself rather than to the additional tool in the agent schema.
We release \sysname{} as an open artifact for reproducibility.\footnote{\url{https://github.com/FARD-Lab/ARISE}}

The paper is organized as follows.
Section~\ref{sec:related} reviews related work.
Section~\ref{sec:methodology} describes the methodology.
Sections~\ref{sec:results} and~\ref{sec:discussion} report and interpret results.
Section~\ref{sec:threats} discusses threats to validity, and Section~\ref{sec:conclusion} concludes.

\section{Related Work}
\label{sec:related}

\paragraph{LLM-based repair and fault localization}
LLM-based APR follows two design philosophies.
\emph{Agentic} systems equip the model with tools and a multi-turn loop to explore, localize, and patch code.
SWE-agent introduces the Agent-Computer Interface for navigating, editing, and testing repositories~\cite{sweagent2024neurips}; AutoCodeRover navigates through structure-aware search APIs before patching iteratively~\cite{zhang2024autocoderover}; and OrcaLoca adds priority-based scheduling and distance-aware context pruning~\cite{yu2025orcaloca}.
\emph{Agentless} systems instead fix a localization, generation, and validation pipeline, matching agents at lower cost~\cite{xia2025agentless}.
On the localization side, traditional spectrum-based and mutation-based methods rank statements from test outcomes~\cite{jones2005tarantula, papadakis2015mutationbasedfl}, while LLM-based methods reason jointly over issue text and code~\cite{liu2026flsurvey, qin2024agentfl, sepidband2026rgfl}.
A consistent finding across this work is that file-level localization is well served by lexical signals, whereas function-level and line-level localization remains the binding constraint on repair~\cite{hossain2024-bug-localization-survey}.
\sysname{} builds on the agentic paradigm and targets this bottleneck with structured graph tools rather than lexical retrieval or free-form navigation.

\paragraph{Graph-based retrieval for code agents}
Several systems expose repository graphs to LLM agents.
RepoGraph builds a line-level reference graph as a SWE-agent plugin~\cite{repograph2025iclr}, LocAgent parses code into heterogeneous file, class, and function graphs with search and traversal tools~\cite{chen2025locagent}, CodexGraph drives a Neo4j database through agent-issued Cypher queries~\cite{liu2025codexgraph}, and KGCompass links issues and pull requests to code entities for path-guided retrieval~\cite{yang2025kgcompass}.
These systems demonstrate the value of structural context, but none models intra-procedural definition-use relationships or exposes a data-flow slice as a first-class agent tool; even RepoGraph, whose reference graph reaches line granularity, captures only lexical reference edges rather than how a value is defined and propagated within a procedure.
Table~\ref{tab:prior_systems} summarizes this comparison.
A parallel line of retrieval-augmented methods for code, including RepoCoder~\cite{zhang2023repocoder}, RLCoder~\cite{wang2025rlcoder}, and GraphCoder~\cite{graphcoder2024ase}, improves completion by retrieving relevant context, but it targets generation and does not expose an interactive analysis primitive that a repair agent can query mid-reasoning.
\sysname{} closes the gap by extending the graph to statement granularity and adding intra-procedural def-use edges, so the \getslice{} tool answers ``which statements affect this variable?'' in a single query, and it packages this analysis as a host-agnostic toolset with a defined integration contract rather than as a host-specific component.

\begin{table*}[t]
\centering
\caption{Comparison of graph-based APR and localization systems.
``Slice API'' denotes whether a data-flow slice can be queried as a structured agent tool.
ARISE is the only system that extends the graph to statement granularity and exposes data-flow slicing as a first-class agent tool.}
\label{tab:prior_systems}
\small
\begin{tabular}{lllcc}
\toprule
\textbf{System} & \textbf{Graph granularity} & \textbf{Data-flow edges} &
  \textbf{Agent tool API} & \textbf{Slice API} \\
\midrule
KGCompass~\cite{yang2025kgcompass}   & File / function     & None (issue/PR metadata) & Yes (GPT-4)  & \ding{55} \\
RepoGraph~\cite{repograph2025iclr}   & Line-level          & Reference edges only      & Plugin       & \ding{55} \\
LocAgent~\cite{chen2025locagent}    & File / class / fn   & None                      & Yes          & \ding{55} \\
CodexGraph~\cite{liu2025codexgraph}  & Module / class / fn & None                      & LLM queries  & \ding{55} \\
\midrule
\sysname{} (ours)   & Package $\to$ statement & AST def-use edges & Yes      & \checkmark \\
\bottomrule
\end{tabular}
\end{table*}

\paragraph{Program slicing}
Program slicing reduces a program to the statements relevant to a slicing criterion, using backward slicing to find statements that influence a variable and forward slicing to find those it influences, typically over a program dependence graph~\cite{weiser1981program, horwitz1990interprocedural, tipp1995slicing}.
Despite its long history, slicing has seen little integration with LLM agents, where program analysis is usually a preprocessing step rather than an interactive, queryable tool.
\sysname{} implements intra-procedural backward and forward slicing over def-use edges and exposes the result through \getslice{}, turning analysis into an on-demand reasoning primitive.

\section{Methodology}
\label{sec:methodology}

\sysname{} comprises three phases (Figure~\ref{fig:overview}).
\emph{Phase~1} constructs a multi-granularity program graph that augments structural relationships with intra-procedural data-flow edges.
\emph{Phase~2} exposes the graph through a three-tier tool API, layering structural navigation, data-flow slicing, and context bundling.
\emph{Phase~3} is the host agent loop, in which an LLM-based agent queries these tools through its host framework to produce either a ranked list of fault locations or a unified diff.
Phases~1 and~2 are the portable \sysname{} toolset, while Phase~3 is supplied by whichever host framework mounts it.
The host's own built-in tools are kept fixed in all experiments, so every reported difference is attributable to the \sysname{} tool layer.

\subsection{Problem Formulation}
\label{sec:problem}

Given a natural-language issue description $I$ and a repository snapshot $\mathcal{R}$ at a fixed commit, we evaluate FL and APR independently, each with its own system prompt and output format. Running the tasks separately lets us isolate the contribution of the \sysname{} tools to localization quality from their contribution to repair success (Section~\ref{sec:eval_protocol}).

\begin{description}[leftmargin=1.5em, noitemsep]
  \item[\emph{Localization.}] The agent returns a ranked list $\mathcal{L} = [(f_i, g_i, \ell_i)]_{i=1}^{k}$, where $f_i$ is a file path, $g_i$ an enclosing function or method, and $\ell_i$ a line number.
  \item[\emph{Repair.}] The agent returns a unified diff $\hat{p}$ such that applying $\hat{p}$ to $\mathcal{R}$ causes all instance-specific tests to pass.
\end{description}

\subsection{Multi-Granularity Repository Graph}
\label{sec:graph}

\sysname{} represents a Python repository as a directed, typed property graph $G = (V, E)$ that unifies \emph{structural} relationships, describing how the codebase is organised into packages, modules, classes, functions, and statements, with \emph{data-flow} relationships, describing how variable values are defined and propagated within functions.
Each node carries a type from $\{$\textsc{Directory}, \textsc{Module}, \textsc{Class}, \textsc{Function}, \textsc{Method}, \textsc{Statement}$\}$ with source-location attributes (\texttt{file\_path}, \texttt{start\_line}, \texttt{end\_line}), and each edge a type from $\{$\textsc{Contains}, \textsc{Imports}, \textsc{ImportedBy}, \textsc{Calls}, \textsc{CalledBy}, \textsc{Inherits}, \textsc{DataflowDefUse}, \textsc{DataflowUseDef}$\}$.
Construction proceeds in two passes.

\paragraph{Structural pass}
We parse every \texttt{.py} file with Python's \texttt{ast} module, emitting \textsc{Directory}, \textsc{Module}, \textsc{Class}, \textsc{Function}, and \textsc{Method} nodes with \textsc{Contains}, \textsc{Imports}/\textsc{ImportedBy}, and \textsc{Inherits} edges.
Inter-module \textsc{Calls}/\textsc{CalledBy} edges require resolving each call site against the import-alias map of the caller's module, which a second pass performs.
Spurious call edges lead agents down incorrect paths and are more harmful than missing edges~\cite{utture2022striking, shi2023distracted}, so we resolve only unambiguous direct and qualified \texttt{module.function()} calls and drop dynamic dispatch through attribute access.

\paragraph{Program graph pass}
This pass extracts data-flow. For every \textsc{Function} and \textsc{Method}, we emit one \textsc{Statement} node per top-level AST statement, connected to the enclosing function by a \textsc{Contains} edge.
\textsc{Statement} nodes are the endpoints of all data-flow edges, so each \textsc{DataflowDefUse}/\textsc{DataflowUseDef} edge connects a defining statement to a using statement rather than individual name occurrences.
We then run an intra-procedural reaching-definition scan.
Definitions are statements where a variable receives a value (\texttt{ast.Assign}, \texttt{ast.AugAssign}, \texttt{ast.AnnAssign}, loop and context-manager targets, and function parameters), and uses are \texttt{ast.Name} nodes in a \texttt{Load} context.
For each use of variable $v$ at statement $s$, we find the last preceding definition of $v$ in source order within the same function, respecting lexical scope (\texttt{global}/\texttt{nonlocal} handled explicitly), and emit a directed \textsc{DataflowDefUse} edge to $s$ with its \textsc{DataflowUseDef} reverse.
A backward slice from $v$ at $s$ follows \textsc{DataflowUseDef} edges to where $v$ was defined; a forward slice follows \textsc{DataflowDefUse} edges to where $v$ is consumed.
The graph is built once per repository and cached, so construction cost is amortized across the instances that share a repository (Section~\ref{sec:cost}).

\subsection{Agentic Tool API}
\label{sec:tools}

\sysname{} exposes $G$ as a set of typed, JSON-schema-annotated functions the backbone can invoke in any order.
A session typically proceeds through navigation, then analysis, then assembly, but no order is imposed.
These stages map to three tiers, separated so each can be added to the tool schema independently for ablation (Section~\ref{sec:eval_protocol}).
All tools share a \texttt{RetrievalSession} holding $G$, a TF-IDF entity index, and an enclosing-scope index, built once per session.

\paragraph{Tier~1: structural retrieval}
Five tools operate on the structural portion of $G$ and reproduce the capabilities of prior structural-graph systems~\cite{repograph2025iclr, yang2025kgcompass, yu2025orcaloca, chen2025locagent}, constituting the \condstruct{} baseline:
\searchentities{} (TF-IDF search over entity names, paths, and docstrings),
\traverserelations{} (bounded breadth-first traversal along given edge types),
\getenclosing{} (maps a \texttt{(file, line)} pair to enclosing \textsc{Function}/\textsc{Class}/\textsc{Module} nodes),
\getcodespan{} (raw source for a line range), and
\getentityinfo{} (metadata summary for a node).

\paragraph{Tier~2: data-flow slicing}
\getslice{} traces a named variable through its enclosing function via the def-use edges in $G$.
Given a seed \texttt{(file, line, variable)} and a direction (\texttt{backward}, \texttt{forward}, or \texttt{both}), it maps the seed to a \textsc{Statement} node, runs a bounded BFS over the appropriate data-flow edges, stops at function boundaries, and returns an ordered list of \texttt{SliceStep} records carrying the file, line range, variable, and role (\emph{parameter}, \emph{definition}, \emph{augmented assignment}, \emph{use}).
If the seed line has no \textsc{Statement} node, \getslice{} returns an empty slice and the agent falls back to \getcodespan{}.

\paragraph{Tier~3: context bundling and ranking}
\buildcontext{} assembles a ranked set of code spans under a token budget (default 8\,000).
Given seed entities $\mathcal{S}$ and optional slices $\mathcal{D}$, each candidate span $c$ is scored as
\begin{equation}
  \text{score}(c) = \alpha \cdot \text{rel}(c)
                  + \beta  \cdot \text{prox}(c)
                  + \gamma \cdot \mathbf{1}[c \in \mathcal{D}],
  \label{eq:bundle_score}
\end{equation}
where $\text{rel}(c)$ is TF-IDF relevance to the issue, $\text{prox}(c)$ is inverse hop distance to the nearest seed in the \textsc{Calls}/\textsc{Imports} subgraph, and the indicator marks slice membership.
The weights $\alpha, \beta, \gamma$ are fixed heuristic constants set to $1.0$, $0.5$, and $1.5$, up-weighting slice membership, and spans are greedily packed in descending score until the budget is exhausted.
\ranksuspect{} ranks suspicious functions from the issue text and any stack trace, expanding seeds through the \textsc{Calls}/\textsc{CalledBy} subgraph and scoring them by the same combination.
It serves as a structural surrogate for inter-procedural data-flow, bringing a callee into the suspect list when the root cause lies past a call boundary the slice cannot cross.

\subsection{Agent Loop}
\label{sec:agent}

The \sysname{} toolset requires only that the host framework expose a tool-call interface, namely the ability to inject the \sysname{} tool schema into the model context and to dispatch the resulting tool calls across a session, a contract any framework can meet (Section~\ref{sec:generalization}).
For the controlled experiments the host is SWE-agent~\cite{sweagent2024neurips}, whose agent-computer interface, multi-turn loop, and patch-generation sub-agent we reuse unchanged; \sysname{} adds only the program-graph tool layer injected into the tool schema at every turn.

The two tasks (APR and FL) use distinct system prompts and output formats, as described in Section \ref{sec:problem}.
In both, the prompt is extended only with the \sysname{} tool-schema descriptions, and the tool docstrings are the sole description of each tool's semantics.
Algorithm~\ref{alg:agent_loop} gives the host agent loop, instantiated with SWE-agent's protocol; at each turn the backbone receives the full history and either emits a tool call (dispatched to a host built-in or a \sysname{} function) or a terminal answer, and the loop ends on a terminal answer or after $T_{\max}$ turns.

\begin{algorithm}[t]
\caption{Host agent loop, instantiated with SWE-agent in our main experiments.
$\mathcal{T} = \mathcal{T}_{\text{host}} \cup \mathcal{T}_{\text{ARISE}}$ combines the host's built-in tools with the portable \sysname{} graph tools (Section~\ref{sec:tools}).
The prompt $P$ and answer schema differ by task, returning a ranked list of \texttt{(file, fn, line)} triples (localization) or a unified diff~$\hat{p}$ (repair).}
\label{alg:agent_loop}
\begin{algorithmic}[1]
\Require Issue $I$, task prompt $P$, session $\mathcal{G}$, tool set $\mathcal{T}$,
         max turns $T_{\max}$
\State $H \leftarrow [P,\, \text{user: } I]$
\For{$t = 1, \ldots, T_{\max}$}
  \State $r \leftarrow \text{LLM}(H,\, \mathcal{T})$
  \If{$r$ is terminal answer}
    \State \textbf{break}
  \EndIf
  \State $\text{result} \leftarrow \mathcal{T}.\text{dispatch}(r.\text{tool\_call})$
  \Comment{host or \sysname{} tool}
  \State $H \leftarrow H \,\|\, [r,\, \text{tool\_result: result}]$
\EndFor
\If{no terminal answer produced}
  \State $r \leftarrow \text{LLM}(H \,\|\, [\text{``produce final answer now''}],\, \mathcal{T})$
\EndIf
\State \textbf{return} $\text{ParseAnswer}(r, P)$
\Comment{ranked location list \emph{or} unified diff}
\end{algorithmic}
\end{algorithm}

\paragraph{Tool availability by condition}
The ablation conditions differ only in the tools exposed (defined in Section~\ref{sec:setup}).
\condstruct{} exposes Tier~1 only, directly comparable to LocAgent~\cite{chen2025locagent} and RepoGraph~\cite{repograph2025iclr}; \condslice{} adds Tier~2; \condfull{} adds all three tiers.
\condcoarse{} exposes the Tier~1 and Tier~2 schemas but builds the graph without \textsc{Statement} nodes or \textsc{Dataflow} edges, so \getslice{} always returns empty, separating the tool API from the data-flow graph.
The prompt structure, turn count, and token budget are held constant across conditions.

\subsection{Evaluation Protocol}
\label{sec:eval_protocol}

\paragraph{Localization metrics}
We score the ranked \texttt{(file, function, line)} list against the gold patch at three granularities.
At \emph{file} level we report File Recall@$k$ (R@$k$, for $k\in\{1,3\}$), the fraction of instances where a gold file appears in the top $k$, and File Mean Reciprocal Rank (MRR).
At \emph{function} level the gold set maps each diff hunk to its enclosing function via \getenclosing{}, and we report Function R@$k$ ($k\in\{1,3\}$), Function MRR, and Function F1@3.
At \emph{line} level the gold set is the union of lines touched by the patch, and we report Line R@$k$ ($k\in\{1,5,10\}$), Coverage@budget (abbreviated Cov; the fraction of gold lines in the \buildcontext{} output under the 8\,000-token budget), and IoU (mean intersection-over-union of predicted and gold line sets).
Coverage@budget is reported only for the \sysname{} conditions, which assemble a budgeted context bundle, and is N/A for the baselines.

\paragraph{Repair metric}
Pass@1 is the fraction of instances whose first generated patch passes all instance tests.
We also report the Spearman rank correlation between Function R@1 and Pass@1, testing whether localization quality correlates with repair success.
Because localization and repair are run as separate sessions, this correlation associates two independent runs on the same instances rather than tracing localization inside the repair run (Section~\ref{sec:threats}).

\section{Results}
\label{sec:results}

\subsection{Experimental Setup}
\label{sec:setup}

\paragraph{Benchmark}
We evaluate on \bench{}~\cite{jimenez2023swe}, comprising 300 real GitHub issues from 11 Python open-source repositories.
Each instance pairs a natural-language issue with a gold patch and a project-specific test harness.
\bench{} is a standard evaluation suite for recent repository-level systems~\cite{repograph2025iclr, yang2025kgcompass, mu2025experepair, xia2025agentless, yu2025orcaloca, zhang2024autocoderover, ma2025lingmaagent, amazon2024qdeveloper}.

\paragraph{Backbone model}
The backbone for all main experiments is \backbone{}~\cite{hui2024qwen25coder}, served locally with vLLM~\cite{kwon2023vllm}.
The maximum agent turn count is 40 and the \buildcontext{} token budget is 8\,000 tokens.

\paragraph{Reproducibility}
All backbones are open-weight and served through vLLM with fixed decoding settings, and evaluation uses the standard \bench{} Docker harness, so every condition is reproducible from the released graph builder, tool API, and run scripts.

\paragraph{Ablation conditions}
The two baselines are \condtext{} (BM25 over raw files, no graph or agent) and the unmodified SWE-agent (built-in tools, no graph).
The \sysname{} conditions add tools cumulatively, namely \condstruct{} (Tier~1, matching LocAgent and RepoGraph), \condslice{} (Tier~1 and 2, the primary novelty condition), and \condfull{} (all three tiers), together with a control condition \condcoarse{} (the Tier~1 and 2 schemas but no \textsc{Statement} nodes, isolating the tool API from the graph); the tools per tier are defined in Section~\ref{sec:tools}.
All \sysname{} conditions share the same host, prompt, turn count, backbone, and harness, so the only variable is the set of tools exposed to the agent.

\paragraph{Baselines}
We select baselines across three tiers. \condtext{} is a lexical lower bound using BM25 file retrieval with no agent loop, establishing the floor from which agentic and graph approaches must improve. SWE-agent~\cite{sweagent2024neurips} and AutoCodeRover~\cite{zhang2024autocoderover} represent SOTA agentic frameworks, both among the top open-submission systems on the SWE-bench leaderboard and the most widely reproduced agentic baselines in recent APR literature. RepoGraph~\cite{repograph2025iclr}, and KGCompass~\cite{yang2025kgcompass} are the most closely related graph-augmented systems; OrcaLoca~\cite{yu2025orcaloca} and Agentless~\cite{xia2025agentless} cover a SOTA localization-first and a non-agentic pipeline respectively.

\subsection{RQ1: Effect of \sysname{} on Bug Repair}
\label{sec:rq1_repair}
\label{sec:repair_results}

\begin{table}[t]
\centering
\caption{End-to-end repair on \bench{}.
Resolved is the instance count resolved out of 300.
Avg.\ tok.\ is per-instance total (input plus output, all turns, $\times1000$).
$\rho$ is the Spearman correlation between Function R@1 and Pass@1.
\texttt{---} marks a value we do not report for that system.
Rows marked * are adopted from~\cite{repograph2025iclr}.
Shading: per-column \colorbox[rgb]{0.957,0.647,0.510}{lowest} to \colorbox[rgb]{0.573,0.773,0.871}{highest}.}
\label{tab:repair}
\small
\setlength{\tabcolsep}{3pt}
\resizebox{\columnwidth}{!}{%
\begin{tabular}{lcccc}
\toprule
\textbf{Method}
  & \textbf{Pass@1} & \textbf{Resolved}
  & \textbf{Avg.\ tok.} & $\boldsymbol{\rho}$ \\
\midrule
\multicolumn{5}{l}{\textit{Baselines (\backbone{})}} \\[2pt]
\condtext{}              &\cellcolor[rgb]{0.957,0.647,0.510} 2.67\%  &\cellcolor[rgb]{0.957,0.647,0.510}  8 &  13  & 0.05 \\
SWE-agent                &\cellcolor[rgb]{0.842,0.685,0.618} 17.3\%  &\cellcolor[rgb]{0.842,0.685,0.618} 52 & 510  & 0.38 \\
SWE-agent+RepoGraph      &\cellcolor[rgb]{0.726,0.722,0.726} 19.3\%  &\cellcolor[rgb]{0.726,0.722,0.726} 58 & ---  & ---  \\
Agentless                &\cellcolor[rgb]{0.918,0.660,0.546} 13.3\%  &\cellcolor[rgb]{0.918,0.660,0.546} 40 & ---  & ---  \\
KGCompass                &\cellcolor[rgb]{0.880,0.672,0.582} 15.3\%  &\cellcolor[rgb]{0.880,0.672,0.582} 46 & ---  & ---  \\
\midrule
\multicolumn{5}{l}{\textit{\sysname{} conditions (\backbone{})}} \\[2pt]
\condstruct{}   &\cellcolor[rgb]{0.765,0.710,0.690} 19.0\%  &\cellcolor[rgb]{0.765,0.710,0.690} 57 & 531  & 0.42 \\
\condcoarse{}   &\cellcolor[rgb]{0.688,0.735,0.762} 19.5\%  &\cellcolor[rgb]{0.688,0.735,0.762} 59 & 535  & 0.43 \\
\condslice{}    &\cellcolor[rgb]{0.611,0.760,0.835} 21.0\%  &\cellcolor[rgb]{0.611,0.760,0.835} 63 & 550  & 0.51 \\
\condfull{}     &\cellcolor[rgb]{0.573,0.773,0.871} \textbf{22.0\%} &\cellcolor[rgb]{0.573,0.773,0.871} \textbf{66} & 560 & \textbf{0.53} \\
\midrule
\multicolumn{5}{l}{\textit{Literature (GPT-4o)}} \\[2pt]
SWE-agent*           &\cellcolor[rgb]{0.803,0.697,0.654} 18.3\%  &\cellcolor[rgb]{0.803,0.697,0.654} 55 & 498  & --- \\
SWE-agent+RepoGraph* &\cellcolor[rgb]{0.649,0.747,0.798} 20.3\%  &\cellcolor[rgb]{0.649,0.747,0.798} 61 & 519  & --- \\
\bottomrule
\end{tabular}%
}
\end{table}

As Table~\ref{tab:repair} shows, \condfull{} resolves 22.0\% (66/300).
The central comparison is the within-setting delta over the SWE-agent host, since both run with the same backbone, prompt, turn budget, and harness, so the toolset is the only changed variable.
On that comparison \condfull{} raises Pass@1 from 17.3\% to 22.0\%, a gain of 4.7 points and 14 additional resolved instances, a relative improvement of roughly 27\%.

Absolute resolve rates are bounded by the open-weight \backbone{} backbone rather than by the toolset, so our claim is not a leaderboard ranking against systems built on proprietary models but the consistent improvement that \sysname{} adds within a fixed setting.
Under that lens \condfull{} outperforms every same-backbone baseline, including the graph-augmented SWE-agent+RepoGraph~\cite{repograph2025iclr} (19.3\%) and the localization-first and non-agentic systems Agentless~\cite{xia2025agentless} (13.3\%) and KGCompass~\cite{yang2025kgcompass} (15.3\%).
It also surpasses the published SWE-agent+RepoGraph result that uses the proprietary GPT-4o model (20.3\%)~\cite{repograph2025iclr}, showing that a graph-augmented open-source configuration can outperform a proprietary-backbone one.
This repair gain comes at a modest token premium over the bare host, which we analyze in Section~\ref{sec:cost}.

\textbf{Finding \#1.} With an open-source 32B backbone, \sysname{} raises Pass@1 over its SWE-agent host by 4.7 points (17.3\% to 22.0\%, a 27\% relative gain), outperforms every same-backbone baseline, and exceeds the RepoGraph+SWE-agent configuration reported with GPT-4o.

\subsection{RQ2: Localization as the Mechanism}
\label{sec:rq2_localization}
\label{sec:localization_results}

We test whether the repair gain is explained by improved fault localization.
Table~\ref{tab:localization} reports localization at file, function, and line granularity against the gold patch.

\begin{table*}[t]
\centering
\caption{\bench{} localization at file, function, and line granularities. Cov is N/A for baselines, applying only to \sysname{}'s budgeted context bundles. \textbf{Bold}: best; \underline{underlined}: best baseline. Shading: per-column \colorbox[rgb]{0.957,0.647,0.510}{lowest} to \colorbox[rgb]{0.573,0.773,0.871}{highest}.}
\label{tab:localization}
\small
\setlength{\tabcolsep}{3pt}
\begin{tabular}{lcccccccccccc}
\toprule
 & \multicolumn{3}{c}{\textbf{File}}
 & \multicolumn{4}{c}{\textbf{Function}}
 & \multicolumn{5}{c}{\textbf{Line}} \\
\cmidrule(lr){2-4}\cmidrule(lr){5-8}\cmidrule(lr){9-13}
\textbf{Condition} & R@1 & R@3 & MRR & R@1 & R@3 & F1@3 & MRR & R@1 & R@5 & R@10 & Cov & IoU \\
\midrule
\multicolumn{13}{l}{\textit{Baselines}} \\[2pt]
\condtext{}   &\cellcolor[rgb]{0.957,0.647,0.510}0.30 &\cellcolor[rgb]{0.957,0.647,0.510}0.42 &\cellcolor[rgb]{0.957,0.647,0.510}0.33 &\cellcolor[rgb]{0.957,0.647,0.510}0.15 &\cellcolor[rgb]{0.957,0.647,0.510}0.25 &\cellcolor[rgb]{0.957,0.647,0.510}0.14 &\cellcolor[rgb]{0.957,0.647,0.510}0.18 &\cellcolor[rgb]{0.957,0.647,0.510}0.05 &\cellcolor[rgb]{0.957,0.647,0.510}0.12 &\cellcolor[rgb]{0.957,0.647,0.510}0.18 & N/A &\cellcolor[rgb]{0.893,0.668,0.570}0.06 \\
SWE-agent     &\cellcolor[rgb]{0.880,0.672,0.582}0.57 &\cellcolor[rgb]{0.803,0.697,0.654}\underline{0.72} &\cellcolor[rgb]{0.880,0.672,0.582}0.60 &\cellcolor[rgb]{0.829,0.689,0.630}\underline{0.43} &\cellcolor[rgb]{0.803,0.697,0.654}\underline{0.58} &\cellcolor[rgb]{0.829,0.689,0.630}\underline{0.39} &\cellcolor[rgb]{0.803,0.697,0.654}\underline{0.48} &\cellcolor[rgb]{0.893,0.668,0.570}0.26 &\cellcolor[rgb]{0.829,0.689,0.630}\underline{0.43} &\cellcolor[rgb]{0.829,0.689,0.630}\underline{0.53} & N/A &\cellcolor[rgb]{0.829,0.689,0.630}\underline{0.21} \\
OrcaLoca      &\cellcolor[rgb]{0.803,0.697,0.654}\underline{0.61} &\cellcolor[rgb]{0.880,0.672,0.582}0.64 &\cellcolor[rgb]{0.803,0.697,0.654}\underline{0.62} &\cellcolor[rgb]{0.893,0.668,0.570}0.35 &\cellcolor[rgb]{0.880,0.672,0.582}0.42 &\cellcolor[rgb]{0.893,0.668,0.570}0.29 &\cellcolor[rgb]{0.880,0.672,0.582}0.43 &\cellcolor[rgb]{0.829,0.689,0.630}\underline{0.28} &\cellcolor[rgb]{0.893,0.668,0.570}0.35 &\cellcolor[rgb]{0.893,0.668,0.570}0.35 & N/A &\cellcolor[rgb]{0.957,0.647,0.510}0.04 \\
\midrule
\multicolumn{13}{l}{\textit{\sysname{} conditions}} \\[2pt]
\condstruct{} &\cellcolor[rgb]{0.726,0.722,0.726}0.62 &\cellcolor[rgb]{0.726,0.722,0.726}0.77 &\cellcolor[rgb]{0.726,0.722,0.726}0.65 &\cellcolor[rgb]{0.765,0.710,0.690}0.50 &\cellcolor[rgb]{0.726,0.722,0.726}0.65 &\cellcolor[rgb]{0.765,0.710,0.690}0.45 &\cellcolor[rgb]{0.726,0.722,0.726}0.54 &\cellcolor[rgb]{0.765,0.710,0.690}0.31 &\cellcolor[rgb]{0.765,0.710,0.690}0.48 &\cellcolor[rgb]{0.765,0.710,0.690}0.59 &\cellcolor[rgb]{0.957,0.647,0.510}0.61 &\cellcolor[rgb]{0.765,0.710,0.690}0.26 \\
\condcoarse{} &\cellcolor[rgb]{0.726,0.722,0.726}0.62 &\cellcolor[rgb]{0.726,0.722,0.726}0.77 &\cellcolor[rgb]{0.726,0.722,0.726}0.65 &\cellcolor[rgb]{0.701,0.731,0.750}0.51 &\cellcolor[rgb]{0.726,0.722,0.726}0.65 &\cellcolor[rgb]{0.701,0.731,0.750}0.46 &\cellcolor[rgb]{0.726,0.722,0.726}0.54 &\cellcolor[rgb]{0.701,0.731,0.750}0.32 &\cellcolor[rgb]{0.701,0.731,0.750}0.49 &\cellcolor[rgb]{0.701,0.731,0.750}0.60 &\cellcolor[rgb]{0.829,0.689,0.630}0.62 &\cellcolor[rgb]{0.701,0.731,0.750}0.27 \\
\condslice{}  &\cellcolor[rgb]{0.649,0.747,0.798}0.65 &\cellcolor[rgb]{0.649,0.747,0.798}0.80 &\cellcolor[rgb]{0.649,0.747,0.798}0.68 &\cellcolor[rgb]{0.637,0.752,0.810}0.57 &\cellcolor[rgb]{0.649,0.747,0.798}0.72 &\cellcolor[rgb]{0.637,0.752,0.810}0.52 &\cellcolor[rgb]{0.649,0.747,0.798}0.60 &\cellcolor[rgb]{0.637,0.752,0.810}0.38 &\cellcolor[rgb]{0.637,0.752,0.810}0.59 &\cellcolor[rgb]{0.637,0.752,0.810}0.71 &\cellcolor[rgb]{0.701,0.731,0.750}0.71 &\cellcolor[rgb]{0.637,0.752,0.810}0.33 \\
\condfull{}   &\cellcolor[rgb]{0.573,0.773,0.871}\textbf{0.67} &\cellcolor[rgb]{0.573,0.773,0.871}\textbf{0.82} &\cellcolor[rgb]{0.573,0.773,0.871}\textbf{0.70}
              &\cellcolor[rgb]{0.573,0.773,0.871}\textbf{0.60} &\cellcolor[rgb]{0.573,0.773,0.871}\textbf{0.75} &\cellcolor[rgb]{0.573,0.773,0.871}\textbf{0.55} &\cellcolor[rgb]{0.573,0.773,0.871}\textbf{0.63}
              &\cellcolor[rgb]{0.573,0.773,0.871}\textbf{0.41} &\cellcolor[rgb]{0.573,0.773,0.871}\textbf{0.62} &\cellcolor[rgb]{0.573,0.773,0.871}\textbf{0.74} &\cellcolor[rgb]{0.573,0.773,0.871}\textbf{0.75} &\cellcolor[rgb]{0.573,0.773,0.871}\textbf{0.36} \\
\bottomrule
\end{tabular}
\end{table*}

\sysname{} achieves the highest score on every metric, outperforming not only the SWE-agent baseline but also OrcaLoca, a purpose-built localization system, at file, function, and line level.
Even the lowest \sysname{} condition (\condstruct{}) outperforms the SWE-agent baseline in R@1 by 0.05 at file level, 0.07 at function level, and 0.05 at line level.
The best condition, \condfull{}, improves R@1 over SWE-agent from 0.57 to 0.67 at file level, 0.43 to 0.60 at function level, and 0.26 to 0.41 at line level, relative gains of 18\%, 40\%, and 58\%, respectively.
The gains are largest at the function and line levels.
File identification is already well served by simple lexical matching, so there is little room to improve, whereas pinpointing the correct function and line needs the structural and data-flow evidence that only the \sysname{} tools provide.
The ranking metrics move the same way.
Function MRR rises from 0.48 to 0.63 and Function F1@3 from 0.39 to 0.55, so \sysname{} not only finds the faulty function more often but also ranks it higher and surfaces fewer wrong candidates.
Coverage@budget, the fraction of gold lines that fit within the 8\,000-token context budget, rises across the \sysname{} conditions from 0.61 (\condstruct{}) to 0.75 (\condfull{}).
This shows that richer tools place more of the faulty region within the agent's working budget, not merely rank it well.
It is N/A for the baselines, which do not assemble a budgeted context bundle.

OrcaLoca, a dedicated localization system, sharpens this picture.
It is the strongest baseline at file level (File R@1 0.61, File MRR 0.62), confirming that file identification is not where the difficulty lies.
At function and line level it falls behind even the SWE-agent baseline (Function R@1 0.35 versus 0.43, Line R@1 0.28), and \condfull{} improves on it by 0.25 in Function R@1, a 71\% relative gain.
OrcaLoca's recall also barely grows with the cutoff (File Recall 0.61 to 0.64 from $k=1$ to $3$, Line Recall flat at 0.35 from $k=5$ to $10$) and its line overlap is low (IoU 0.04), indicating a short, high-precision candidate list.
For a downstream repair agent this low recall at depth is a liability, since a faulty location absent from the list can never be repaired, which is the regime where \sysname{}'s higher recall at every cutoff matters most.
Following prior fault-localization work~\cite{yu2025orcaloca, chen2025locagent, hossain2024-bug-localization-survey}, we focus on recall.
For localization that feeds a repair agent, missing the faulty location is fatal, while a few extra candidates are cheap, since the agent can discard false positives while patching but can never fix a location it never retrieved.

To test whether better localization explains the repair gains, we compute the Spearman correlation $\rho$ between Function R@1 and Pass@1 for each condition (Table~\ref{tab:repair}).
It rises steadily as the tools improve, from 0.05 (\condtext{}) to 0.38 (SWE-agent), 0.42 (\condstruct{}), and 0.53 (\condfull{}).
The rise from 0.42 to 0.51 when slicing is added (\condstruct{} to \condslice{}) lines up with the largest per-tier repair gain, which fits the view that finding the right function is what drives the fix.
This correlation is suggestive but not conclusive.
We measure localization and repair in separate runs, and easy issues tend to be both easier to localize and easier to fix, so difficulty alone could account for part of the correlation.
We record this as a construct threat in Section~\ref{sec:threats}.

A per-instance view supports the same conclusion and avoids the difficulty confound, since it follows the same instances across the two conditions.
Adding slicing newly localizes some instances to the correct function.
Of those, 28\% also turn from unresolved to resolved, against only 3\% of the instances whose localization does not change.
Almost all of the net repair gain at this step, a net of 6 instances, comes from this newly localized set.

\textbf{Finding \#2.}
\sysname{} improves localization most at the function and line levels (relative gains of 40\% Function R@1 and 58\% Line R@1), surpassing even the dedicated localizer OrcaLoca (Function R@1 0.35), and this improvement is correlated with repair success ($\rho$ up to 0.53), consistent with localization being the primary driver of the repair gain.

\subsection{RQ3: Component Contributions}
\label{sec:rq3_ablations}

This subsection isolates each component through controlled comparisons; Tables~\ref{tab:repair} and~\ref{tab:localization} provide the quantitative basis.

Adding Tier~1 tools raises File R@1 by 0.05, Function R@1 by 0.07, and Pass@1 by 1.7 points.
The SWE-agent-to-\condstruct{} repair gain (+1.7 points) is close to the +2.0 point delta between SWE-agent and SWE-agent + RepoGraph on GPT-4o~\cite{repograph2025iclr}, a consistent estimate of what a structural graph contributes.

Adding Tier~2 yields the single largest within-tier improvement, with Function R@1 up 0.07, Line IoU up 0.07, and Pass@1 up 2.0 points (19.0\% to 21.0\%, 6 instances), because \getslice{} directly answers which statements use a suspicious variable and where it was last defined, a query that otherwise requires multiple \traverserelations{} hops to approximate.

Adding Tier~3 yields a further +1.0 point Pass@1 (21.0\% to 22.0\%, 3 instances) and +0.03 Function R@1.
The total gain from SWE-agent to \condfull{} decomposes as +1.7 (structural graph), +2.0 (slicing), and +1.0 (bundling).

\condcoarse{} exposes the \getslice{} schema but builds the graph without \textsc{Statement} nodes or \textsc{Dataflow} edges, so \getslice{} always returns empty.
Its metrics sit within 0.01 of \condstruct{} (Function R@1 0.51 vs.\ 0.50, Pass@1 19.5\% vs.\ 19.0\%), confirming that the \condslice{} improvement comes from the data-flow graph itself and not from the presence of an extra tool in the schema.

\textbf{Finding \#3.}
No tier is redundant. Each one adds a consistent increment, and the data-flow graph rather than the tool schema drives the slicing gain.

\subsection{Portability Across Host Frameworks and Backbones}
\label{sec:generalization}

RQ1 through RQ3 isolate \sysname{} on a single host framework (SWE-agent) and a single backbone (\backbone{}).
Because \sysname{} is a portable toolset rather than a SWE-agent extension, we now mount it on a second host and drive it with an additional backbone model to test whether it ports as a drop-in component.
We mount \sysname{} on SWE-agent and AutoCodeRover~\cite{zhang2024autocoderover}, the two hosts that satisfy the tool-call contract of Section~\ref{sec:tools}, running each standalone and with \sysname{} added (``+\sysname{}''), with a larger backbone, Qwen3.6-35B-A3B.
We extend the study to one additional host because rerunning the full condition suite on a new framework is computationally expensive.
We select AutoCodeRover because it ranks among the top open-submission systems on the \bench{} leaderboard, is widely reproduced in recent APR work, and exposes the tool-call interface \sysname{} requires.

\begin{table}[t]
\centering
\caption{Portability of \sysname{} across host frameworks on  Qwen3.6-35B-A3B. Pass@1 (resolved/300) on \bench{}.}
\label{tab:generalization}
\small
\setlength{\tabcolsep}{4pt}
\begin{tabular}{lc}
\toprule
\textbf{Framework} & \textbf{Pass@1}\\
\midrule
SWE-agent              & 33.3\% (100/300) \\
SWE-agent+RepoGraph    & 33.3\% (100/300) \\
SWE-agent+\sysname{}   & 34.0\% (102/300) \\
\addlinespace
AutoCodeRover            & 36.3\% (109/300) \\
AutoCodeRover+\sysname{} & 32.0\% (96/300) \\
\bottomrule
\end{tabular}
\end{table}

The same toolset mounts on both SWE-agent and AutoCodeRover without modification, which is the portability claim of contribution~(3), but the effect on Pass@1 is host-dependent, rising on SWE-agent (33.3\% to 34.0\% with Qwen3.6-35B-A3B) yet falling on AutoCodeRover (36.3\% to 32.0\%).

The benefit of external augmentation also shrinks as the backbone strengthens.
On Qwen3.6-35B-A3B, SWE-agent+RepoGraph matches the bare SWE-agent (both 33.3\%) and \sysname{} adds only 0.7 points, against the larger margins both show on \backbone{}.
This is consistent with stronger models internalizing the structural navigation that graph tools provide, so we scope our contribution to the open-source regime where this augmentation matters most and leave a controlled study of representation value at frontier scale to future work.

We attribute the AutoCodeRover drop to integration fit rather than to \sysname{} itself. AutoCodeRover is designed around its own search interface, tool signatures, and prompting strategy, which together let the agent use its native tools effectively. When \sysname{} is substituted without adapting this integration, Pass@1 falls to 3.6\%, indicating that the default dispatch layer cannot exercise \sysname{}'s interface. After modifying AutoCodeRover's tool-calling infrastructure to accommodate \sysname{}'s keyword and optional arguments, while leaving the rest of the agent unchanged, Pass@1 recovers to 32.0\%, so most of the initial degradation stems from the integration layer rather than the toolset. AutoCodeRover with its native tools still outperforms AutoCodeRover+\sysname{}, a gap we attribute to the host being tuned end-to-end around its own tooling ecosystem, including its prompts, tool interfaces, and interaction patterns. This is consistent with the portability claim: the toolset ports cleanly, but its benefit depends on how well the host already exploits structure-aware search.

\textbf{Finding \#4.}
The \sysname{} toolset ports as a drop-in component across host frameworks that satisfy the tool-call contract, and the size and direction of the Pass@1 effect depend on what the host already provides.

\section{Discussion}
\label{sec:discussion}

\subsection{Why Data-flow Slicing Helps}
\label{sec:discussion_dataflow}

The \condstruct{} to \condslice{} gain is mediated by function-level localization (Function R@1 rises from 0.50 to 0.57, the largest within-tier jump), while the file-level gain is only 0.03 because issue reports usually name the module or class, so BM25 and \searchentities{} already locate the file.
The remaining bottleneck is distinguishing the faulty function from its structurally adjacent neighbors, which \getslice{} addresses directly, since a backward slice traces the variable's definition chain to the defining statement, collapsing a multi-hop traversal into a single \textsc{DataflowUseDef} breadth-first search (BFS).

\subsection{Failure Analysis}
\label{sec:error_analysis}

\begin{table}[h]
\centering
\caption{Failure-mode breakdown for \condslice{} ($n = 237$ failed instances) and \condfull{} ($n = 234$). Count is the number of unresolved instances.}
\label{tab:errors}
\small
\begin{tabular}{lcccc}
\toprule
 & \multicolumn{2}{c}{\condslice{}}
 & \multicolumn{2}{c}{\condfull{}} \\
\cmidrule(lr){2-3}\cmidrule(lr){4-5}
\textbf{Failure category} & \textbf{Count} & \textbf{\%} &
                             \textbf{Count} & \textbf{\%} \\
\midrule
Wrong file               & 107 & 45\% & 99 & 42\% \\
Right file, wrong fn     &  59 & 25\% & 58 & 25\% \\
Right fn, failed repair  &  47 & 20\% & 51 & 22\% \\
Incomplete localization  &  24 & 10\% & 26 & 11\% \\
\bottomrule
\end{tabular}
\end{table}

Table~\ref{tab:errors} shows where the remaining failures concentrate.
The dominant mode, wrong file at 45\% of \condslice{} failures (107 of 237), is a direct consequence of the intra-procedural scope limit.
\getslice{} stops at \textsc{Calls} edges, so a bug whose root cause is a value mutated inside a callee is invisible to the backward slice.
\condfull{} reduces this from 45\% to 42\% because \ranksuspect{} expands candidates through the \textsc{Calls} subgraph, a structural surrogate rather than a data-flow solution.
The right-function, failed-repair category (20\%) is orthogonal to localization, since these instances are correctly localized but require a multi-hunk edit or a semantic invariant not visible locally, so better localization cannot raise their ceiling.

\subsection{Cost and Efficiency}
\label{sec:cost}

\begin{table}[h]
\centering
\caption{Mean token consumption ($\times1000$) per instance by tool type. Totals include prompt and generation overhead not itemized.}
\label{tab:cost}
\small
\setlength{\tabcolsep}{3.5pt}
\resizebox{\columnwidth}{!}{%
\begin{tabular}{lccccc}
\toprule
\textbf{Condition} & \textbf{Search} & \textbf{Traverse} &
  \textbf{Slice} & \textbf{Bundle} & \textbf{Total} \\
\midrule
\multicolumn{6}{l}{\textit{Baselines}} \\[2pt]
\condtext{}         & N/A      & N/A      & N/A      & N/A      & 13.0 \\
SWE-agent           & N/A      & N/A      & N/A      & N/A      & 510 \\
\midrule
\multicolumn{6}{l}{\textit{\sysname{} conditions}} \\[2pt]
\condstruct{}       & 5.1 & 24.9 & N/A        & N/A        & 531 \\
\condcoarse{}       & 5.1 & 24.9 & 5.0 & N/A  & 535 \\
\condslice{}        & 5.1 & 21.6 & 23.3   & N/A        & 550 \\
\condfull{}         & 5.1 & 21.6 & 23.3   & 10   & 560 \\
\bottomrule
\end{tabular}%
}
\end{table}

\paragraph{Token cost}
Table~\ref{tab:cost} breaks down mean token consumption by tool type.
\sysname{} is not cheaper than its host; \condfull{} consumes about 560{,}000 tokens per instance against 510{,}000 for the bare SWE-agent baseline, so the toolset buys additional fixes at additional token cost rather than saving tokens.
The \condcoarse{} slice column (5\,000 tokens) is empty-slice response overhead, confirming that a failed \getslice{} call is cheap and does not explain the \condslice{} gain, which instead costs a 3.6\% token increase over \condstruct{} for 6 additional resolved instances, near 3\,200 tokens per additional fix.
At roughly 3\,200 additional tokens per extra resolved instance, the overhead is modest relative to the repair gain, and practitioners who prioritize fix rate over cost efficiency could find the trade-off favorable.

\paragraph{Graph construction cost}
The graph is built once per repository and cached to a pickle, after which every instance that shares the repository loads it essentially instantly.
Across the 11 \bench{} repositories, each graph holds a median of 134{,}000 nodes and 313{,}000 edges (up to 264{,}000 and 780{,}000), of which a median of 94{,}000 are statement-level nodes.
First-build time ranges from 1.7s for the smallest repository (\texttt{requests}) through roughly 2.2 minutes for a median one (\texttt{django}) to roughly 33 minutes for the largest (\texttt{sympy}), with the data-flow pass accounting for 96\% to 99\% of the total.
Because this build is paid once and amortized over the many \bench{} instances that share each repository, its per-instance contribution is small even for \texttt{sympy}.

\subsection{How ARISE Impacts Agent Behavior: A Deeper Look}
\label{sec:arise-repair-precision}

All the analysis in this section draws from agent trajectories recorded during the Qwen3.6-35B-A3B experiments, comparing SWE-agent alone against SWE-agent+\sysname{}.
Despite nearly identical aggregate repair rates, the trajectory-level behavior differs substantially, revealing how \sysname{}'s fault localization reshapes the repair process even when the end result looks similar.

\begin{figure}[!t]
  \centering
  \includegraphics[width=\columnwidth]{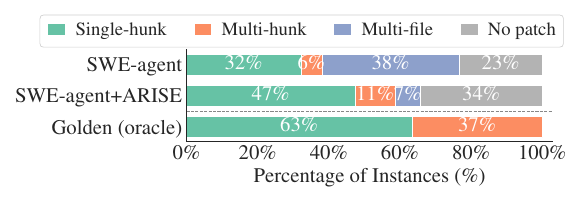}
  \caption{Patch type distribution comparing SWE-agent and \sysname{} with oracle.}
  \label{fig:patch-type}
\end{figure}

\paragraph{Patch Type Distribution}
As Figure~\ref{fig:patch-type} shows, developer-written patches always target a single file.
Without \sysname{}, SWE-agent frequently edits multiple files, spreading changes across unrelated locations.
\sysname{} reduces multi-file patches by 81\%, pulling the distribution close to the oracle.
Multi-file patches are harder to review, more likely to introduce regressions in untouched code, and signal that the agent was uncertain about the fault location; this reduction shows that \sysname{}'s fault localization reduces that uncertainty.

\paragraph{Hunk Count Distribution}
SWE-agent+\sysname{} produces a compact hunk count distribution (median=2, max=3 per patch), closely matching the oracle, while SWE-agent shows a wider spread with outliers reaching 6 hunks.
Fewer hunks mean the agent is editing in one coherent location rather than scattering small changes across a file, which is the structural signature of a more precisely localized patch.

\begin{figure}[!t]
  \centering
  \includegraphics[width=\columnwidth]{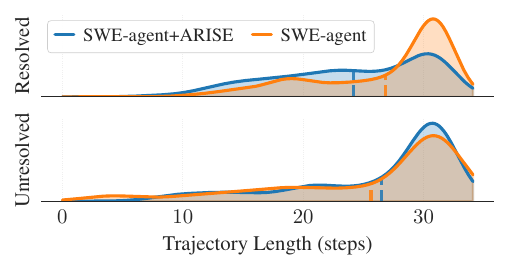}
  \caption{Trajectory length distribution for resolved and unresolved instances across SWE-agent and SWE-agent+\sysname{}.}
  \label{fig:traj-length}
\end{figure}

\paragraph{Trajectory Length Distribution}
As Figure~\ref{fig:traj-length} shows, SWE-agent+\sysname{} resolves instances in fewer steps than SWE-agent.
SWE-agent's step-count distribution for resolved instances shows that it often converges on a fix only as it approaches the 32-step cap.
For unresolved instances, both systems show near identical distributions.

\paragraph{Tool Call Distribution}

Analyzing the mean number of tool calls per problem instance reveals that SWE-agent+\sysname{} makes proportionally more file-edit and test-execution calls while making fewer navigation and environment-setup calls. Among instances that produced a patch, \sysname{} averages 2.52 file-edit calls and 7.22 execution calls per instance, compared to 0.76 and 4.55 for the baseline, a $3.3\times$ and $1.6\times$ increase, respectively. Conversely, navigation calls drop from 17.88 to 12.90 (a $28\%$ reduction) and environment-setup calls drop from 3.02 to 1.22 (a $60\%$ reduction). This shift reflects what fault localization enables; instead of searching broadly for the relevant file, the agent can spend its budget editing and verifying a precise target. In effect, \sysname{} converts navigation time into editing time, the budget allocation that leads to successful patches.

\subsection{Illustrative Example}
\label{sec:illustrative-example}

\begin{figure}[!t]
\centering
\includegraphics[width=\columnwidth]{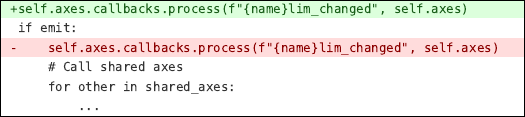}
\caption{The patch diff generated for instance \texttt{matplotlib\_\_matplotlib-26011} using SWE-agent+\sysname{}.}
\label{fig:patch-example}
\end{figure}

We examine instance \texttt{matplotlib\_\_matplotlib \allowbreak-26011} where the baseline produced an empty patch, failing to locate the bug. The fix (Figure~\ref{fig:patch-example}) required understanding a subtle semantic mismatch between the reported symptom and the root cause, a misplaced callback in a private internal method \texttt{\_set\_lim} with no lexical connection. ARISE provides tools such as \texttt{arise\_rank\_suspects} that enabled the agent to bridge this gap. A follow-up tool \texttt{arise\_search} confirmed the exact location in five tool calls before any edit was made.

\subsection{Implications}
\label{sec:implications}

For researchers, the central finding is that data-flow analysis should be treated as a retrieval primitive rather than a preprocessing step. The 0.17 gain in Function R@1 from \sysname{}'s graph and slicing tools, with data-flow slicing the single largest contributor, shows that the precision of what the agent retrieves matters more than the volume, and that structural graphs alone leave a gap that semantic information closes. Systems that invest in richer program representations can close this gap while using smaller, open-source models, which implies that improving graph representation quality is a more cost-effective path to higher repair than scaling up to larger proprietary models.

For practitioners, \sysname{} can be adopted without modifying the host framework: the toolset mounts on any system that exposes a tool-call interface. The benefit depends on what the host already provides structurally, with a host that already has structure-aware retrieval gaining less from the structural tier while still benefiting from the data-flow layer, making the data-flow component the most transferable part of the system.

\section{Threats to Validity}
\label{sec:threats}
\label{sec:limitations}

\subsection{Internal Validity}

The most serious internal threat is benchmark contamination of the backbone.
\bench{} is drawn from public repositories, so its gold patches may appear in \backbone{}'s pretraining data, potentially inflating absolute resolve rates for every condition that uses the model.
This threat is mitigated by the study design: \sysname{}, all baselines, and all ablations share the same backbone, so contamination is held constant and cannot explain the within-setting deltas we report.
We further reduce retrieval-time leakage by building the graph from the base-commit snapshot, preventing future patches and issue comments from entering the agent context; memorization inside the model weights applies equally to all conditions.

Pass@1 is measured from a single patch attempt per instance, so run-to-run sampling variance is not averaged out.
To ensure comparability, we hold decoding configuration, turn count, and token budget constant across all conditions, so observed deltas reflect toolset differences rather than stochastic variation.
Repair success is judged by project test suites, which can pass a patch that edits a different location than the gold patch; this is examined under construct validity.

\subsection{External Validity}

\bench{} covers Python only and favors repositories with well-structured issue reports and deterministic test suites, so the absolute resolve rates may not transfer to other languages or project types.
This is consistent with all recent work evaluated on \bench{}~\cite{repograph2025iclr, yang2025kgcompass, xia2025agentless, yu2025orcaloca, zhang2024autocoderover}, which is the standard evaluation suite for repository-level APR, making our results directly comparable to prior work.

The def-use analysis is Python-specific, using \texttt{ast} node types; extending \sysname{} to other languages would require language-specific frontends.
The intra-procedural analysis covers the most common Python assignment patterns; when a pattern is not instrumented, \getslice{} returns an empty slice and the agent falls back to \getcodespan{}, so the typical effect of a missed pattern is neutral degradation rather than a wrong result.

\subsection{Construct Validity}
Pass@1 measures whether the first patch passes all project tests, which can diverge from editing the ground-truth location.
\bench{}'s project test suites are a widely accepted and validated proxy for repair correctness, used as the primary metric across the recent APR literature~\cite{jimenez2023swe, repograph2025iclr, xia2025agentless}, so the metric measures what we intend to measure.
To further guard against over-reliance on a single metric, we separately report localization recall at file, function, and line granularity and examine the Spearman correlation between localization and repair success; the two are measured in independent sessions so the correlation reflects an association rather than a within-run causal trace.

The most significant architectural limitation is that \getslice{} does not cross \textsc{Calls} edges and so cannot trace a root cause mutated inside a callee.
This failure mode is analyzed as the dominant wrong-file category in Section~\ref{sec:error_analysis} and is the primary direction for future work.
It reflects a scope boundary of the current design rather than a flaw in the experimental methodology.

\section{Conclusions}
\label{sec:conclusion}

We presented \sysname{}, a portable, framework-agnostic toolset that builds a multi-granularity program graph and exposes intra-procedural data-flow slicing as a first-class agent primitive, mounting on any host framework that supports tool calls.
The central finding is that data-flow precision matters more than model scale, since a graph-augmented open-source model outperforms configurations using larger proprietary models because data-flow slicing lets the agent answer ``which statements define this variable?'' in a single call rather than searching broadly. Controlled ablations confirm that the data-flow graph, not the mere presence of an extra tool in the schema, drives the gain.
On \bench{} with SWE-agent as host, \sysname{} resolves 22.0\% of issues (66/300), a 4.7 percentage-point gain over SWE-agent that our analysis attributes to sharper fault localization (Function R@1 rising from 0.43 to 0.60) and, through controlled ablations, to the data-flow graph rather than the tool schema.
\sysname{} is portable across host frameworks, with the benefit depending on what the host already provides.
The primary direction for future work is extending slicing from intra-procedural to inter-procedural scope; a secondary direction is incorporating data-flow representations into RAG-based retrieval for non-agentic repair systems.

{\footnotesize
\bibliography{ref}}

\end{document}